\documentclass[runningheads]{llncs}
\usepackage[T1]{fontenc}

\usepackage{graphicx}
\usepackage{color}
\usepackage{url}
\usepackage{hyperref}
\usepackage{array}
\usepackage{marvosym}
\usepackage{hyperref}
\usepackage{xcolor}

\newcommand{\correspondingauthor}[1]{\href{mailto:#1}{{\textcolor{black}{(\scalebox{1.2}{\Letter})}}}}

\newcolumntype{L}[1]{>{\raggedright\let\newline\\\arraybackslash\hspace{0pt}}m{#1}}
\newcolumntype{C}[1]{>{\centering\let\newline\\\arraybackslash\hspace{0pt}}m{#1}}
\newcolumntype{R}[1]{>{\raggedleft\let\newline\\\arraybackslash\hspace{0pt}}m{#1}}

\begin{document}
\title{RadiomicsFill-Mammo: Synthetic Mammogram Mass Manipulation with Radiomics Features}
\titlerunning{RadiomicsFill-Mammo}
%
\author{
    Inye Na\inst{1} \and
    Jonghun Kim\inst{1}\and
    Eun Sook Ko\inst{2} \and
    Hyunjin Park\inst{1,3}\thanks{Corresponding author}
}

\authorrunning{I. Na et al.}

\institute{
    Department of Electrical and Computer Engineering, \\ Sungkyunkwan University, Suwon 16419, South Korea \\
    \and
    Department of Radiology, Samsung Medical Center, \\ Sungkyunkwan University School of Medicine, Seoul 06351, South Korea
    \and
    Center for Neuroscience Imaging Research, \\ Institute for Basic Science, Suwon 16419, South Korea  \\
    \email{hyunjinp@skku.edu} 
}

\maketitle
\begin{abstract}

Motivated by the question, ``Can we generate tumors with desired attributes?'' this study leverages radiomics features to explore the feasibility of generating synthetic tumor images. Characterized by its low-dimensional yet biologically meaningful markers, radiomics bridges the gap between complex medical imaging data and actionable clinical insights. We present \textbf{\textit{RadiomicsFill-Mammo}}, the first of the \textit{RadiomicsFill} series, an innovative technique that generates realistic mammogram mass images mirroring specific radiomics attributes using masked images and opposite breast images, leveraging a recent stable diffusion model. This approach also allows for the incorporation of essential clinical variables, such as BI-RADS and breast density, alongside radiomics features as conditions for mass generation. Results indicate that \textit{RadiomicsFill-Mammo} effectively generates diverse and realistic tumor images based on various radiomics conditions. Results also demonstrate a significant improvement in mass detection capabilities, leveraging \textit{RadiomicsFill-Mammo} as a strategy to generate simulated samples. Furthermore, \textit{RadiomicsFill-Mammo} not only advances medical imaging research but also opens new avenues for enhancing treatment planning and tumor simulation. Our code is available at \url{https://github.com/nainye/RadiomicsFill}.

\end{abstract}

\section{Introduction}
In medical imaging, acquiring high-quality data is challenging due to privacy concerns and high labeling costs. Additionally, the inherent imbalance in available datasets complicates the training of effective neural networks, presenting a significant barrier to advancements in medical artificial intelligence (AI) \cite{shorten2019survey,gao2020handling}. To overcome these challenges, researchers have explored new methods, including the generation of synthetic data using advanced techniques like Generative Adversarial Networks and diffusion models \cite{shin2018medical,frid2018gan,Kim_2024_WACV,garcea2023data,ozbey2023unsupervised,na2023synthetic}. Rouzrokh et al. \cite{rouzrokh2022multitask} developed a denoising diffusion model for brain MRI slice inpainting, enabling customization with five types of masks, including tumor-free and various tumor states. Sagers et al. \cite{sagers2022improving} utilized DALL·E 2 to generate photorealistic skin disease images for diverse skin types, improving classification accuracy, notably for underrepresented groups. These methods aim to bridge the gaps in data availability and balance but often fall short in customizing synthetic data with specific features.

Radiomics features, despite being lower in dimension than the full images, carry important biological information that has proven useful in diagnosis and prognosis \cite{traverso2018repeatability,na2024combined,ibrahim2021radiomics}. Radiomics is less likely to overfit compared to models trained directly on images, especially when the available data is limited \cite{parmar2014robust,zwanenburg2019assessing}.

Building on this foundation, we introduce \textit{RadiomicsFill-Mammo}, a novel approach that leverages radiomics features for generating synthetic mammogram images. This method can help balance out data and provide more specific control over the generation of realistic tumors. Our approach utilizes stable diffusion \cite{Rombach_2022_CVPR} as the backbone generator for \textit{RadiomicsFill-Mammo}, selected for its denoising-based operation and flexibility in handling various input formats, which is crucial for mammography. The generator's pretraining on large-scale datasets ensures a robust foundation, enhancing the quality and realism of synthetic images. As the first in the \textit{RadiomicsFill} series, \textit{RadiomicsFill-Mammo} demonstrates the practicality and transformative potential of employing radiomics features in medical imaging research and clinical practices. Our main contributions are:

\begin{enumerate}
\item Development of a prompt encoder capable of learning and reflecting specific conditions from tabular radiomics features for accurate tumor generation.
\item Incorporation of clinical conditions and optimization for mammography environments, including images of the opposite breast, breast density, and Breast Imaging Reporting \& Data System (BI-RADS) ratings, enhancing medical applicability.
\item Evidence of \textit{RadiomicsFill-Mammo}'s effectiveness as an augmentor in downstream tasks, particularly in mass detection, underscoring its role in enhancing model performance across medical AI applications.
\item Demonstrated adaptability to small external datasets, showcasing the potential for broad application and research in medical imaging.
\end{enumerate}

\begin{figure}[t]
\includegraphics[width=\textwidth]{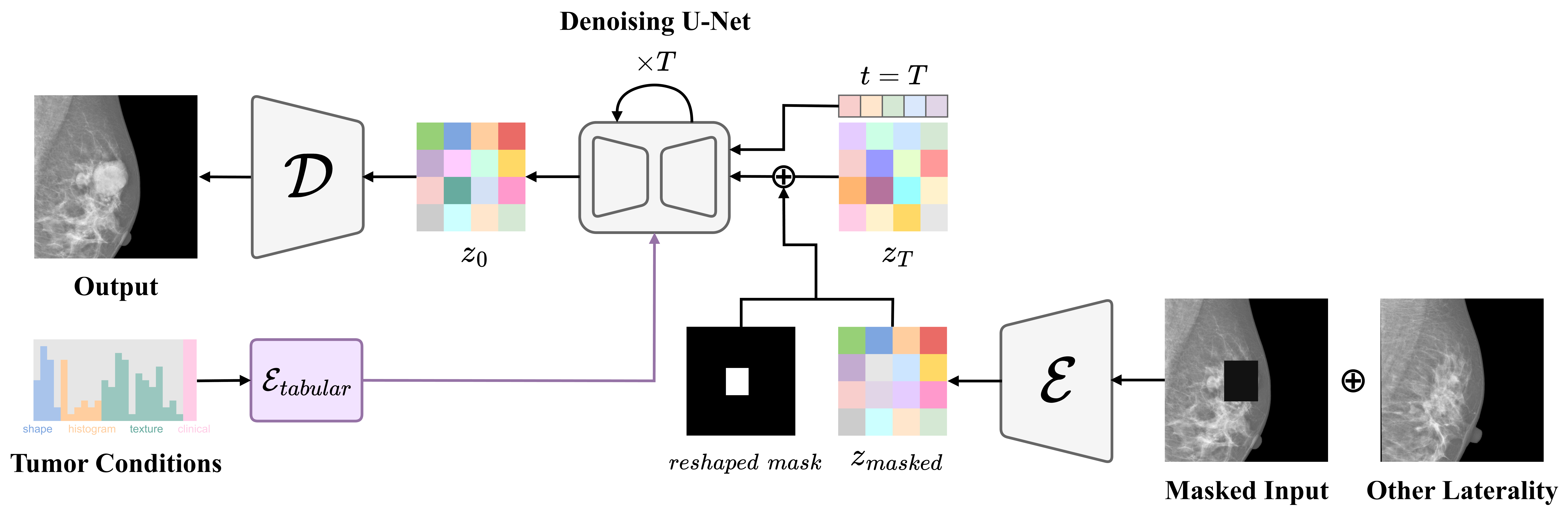}
\caption{\textbf{Overview of the \textit{RadiomicsFill-Mammo} model}, a stable diffusion-based architecture. This model performs iterative denoising for tumor inpainting within masked regions on a noisy latent vector, utilizing information from unmasked regions, the opposite breast, and specific tumor conditions.} \label{fig:main1}
\end{figure}

\section{Method}
\subsection{Dataset}

For our study, we employed the publicly available VinDr-Mammo dataset \cite{nguyen2023vindr} from Vietnam, containing 5,000 full-field digital mammography cases with breast density and annotation details. The data were pre-split into training and test sets at a 4:1 ratio at the patient level. Our focus was on samples with either a Mass finding or No findings. Due to the provision of only bounding boxes for the mass regions, we worked with an experienced breast radiologist to semi-automatically (i.e., vanilla U-Net followed by manual correction) delineate mass contours for extracting radiomics features, employing these in our experiments.

The data utilized in our research was distributed as follows: for mass cases, we allocated 886 masses for training (423 subjects; 26 low density-benign [L-B], 96 low density-malignant [L-M], 376 high density-benign [H-B], and 388 high density-malignant [H-M]), 96 for validation (46 subjects; 4 L-B, 4 L-M, 38 H-B, and 50 H-M), and 235 for testing (115 subjects; 12 L-B, 14 L-M, 110 H-B, and 99 H-M). For normal cases, 12,680 images for training (3,170 subjects), 1,408 for validation (352 subjects), and 3,540 for testing (885 subjects) were used. We trained our model without distinguishing between left/right and cranial-caudal/mediolateral oblique (CC/MLO) mages, incorporating all variations into a single model to leverage the comprehensive dataset for enhanced learning outcomes.

\subsection{RadiomicsFill-Mammo}
\textit{RadiomicsFill-Mammo}, as illustrated in Fig.~\ref{fig:main1}, is a framework designed for inpainting masses in masked regions of mammograms. It takes as input the masked image, an image of the opposite breast (other laterality image), and tumor conditions. The opposite side image is included to leverage breast symmetry, aiding in inpainting by providing a reference \cite{scutt2006breast}. In 2D mammograms, normal and abnormal tissues overlap due to projection, and a rectangular mask may contain both normal and abnormal tissue depending on the radiomics condition. Using the opposite side helps fill the masked region with normal tissue texture while generating masses that reflect the given conditions. 

The generator within this framework is adapted from the stable-diffusion-2-inpainting \cite{Rombach_2022_CVPR} model, with modifications made to suit our specific requirements. To improve tumor generation, we replace the standard text encoder with a pretrained tabular encoder based on radiomics features, as detailed in Section 2.3. This choice tailors the generator's capabilities more closely to medical imaging requirements. We specifically fine-tune the denoising U-Net for mass inpainting, ensuring precise integration of tumor masses into mammograms. This focused adjustment enhances the generated images' realism and clinical relevance.

\subsection{Tabular Encoder}

\begin{figure}[t]
\includegraphics[width=\textwidth]{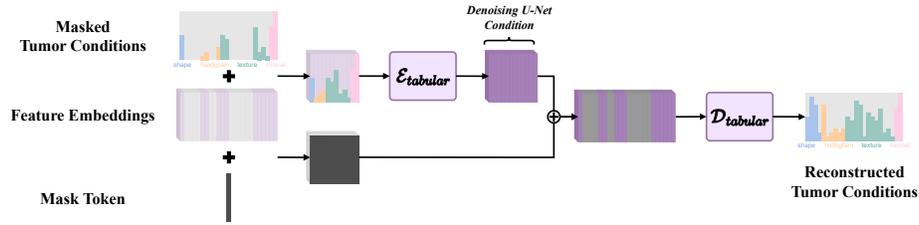}
\caption{ \textbf{Pretraining Process of the Tabular Encoder.} Unmasked feature values (i.e., radiomics features) and their embeddings are concatenated for the encoder input. For masked features, a mask token and the masked feature's embeddings are concatenated, forming the input for the decoder, which focuses on reconstructing tumor conditions. Following pretraining, the encoder is frozen, and its output is utilized for tumor inpainting.} \label{fig:main2}
\end{figure}

We aim to produce prompts detailing tumor characteristics for integration into tumor generation within the stable diffusion framework. A key component of our \textit{RadiomicsFill} model is the tabular encoder, trained as part of the Masked Encoding for Tabular data (MET) \cite{majmundar2022met} process, incorporating its strategy for feature-specific embeddings. This encoder effectively identifies complex feature relationships by reconstructing masked values, acknowledging the stable but variable nature of these characteristics. While some features may be redundant, their relationships are effectively captured by the feature-specific embeddings. The embeddings are structured as \(feature\; number \times embedding\; dimension\), allowing for precise representation of each feature's unique properties. After pretraining on a masked feature reconstruction task with mean squared error loss, this encoder becomes the prompt encoder in our tumor generator. Fig.~\ref{fig:main2} demonstrates the MET network's architecture and its pretraining approach, enabling the generation of clinically relevant and detailed images by leveraging the structured complexity of tabular data.

\subsubsection{Evaluation of the Tabular Encoder}

We adopted a Masked Tumor Condition, incorporating 9 shape, 18 histogram, and 40 texture features, alongside clinical variables (i.e., breast density as low/high and BI-RADS assessments as normal, benign, and malignant). Radiomics features were extracted using PyRadiomics \cite{van2017computational} and normalized to a range of 0 to 1. In the case of normal samples, which lack mass, all radiomics features were set to 0, without any masking in the training process.

To evaluate the pretrained tabular encoder's effectiveness, we employed logistic regression for 
BI-RADS prediction based on radiomics features, with clinical variables masked during evaluation. For  BI-RADS prediction, the baseline model used only the 67 radiomics features, whereas the MET model also included trained feature embeddings, with the encoder output used for logistic regression. This approach allowed for hyperparameter optimization, including the embedding dimension and encoder-decoder architecture, with outcomes detailed in the supplementary materials.

Furthermore, we enhanced learning efficiency by adopting a curriculum learning strategy that progressively increased the feature masking rate.

\subsection{Experimental Setting} 

\begin{figure}[t]
\includegraphics[width=\textwidth]{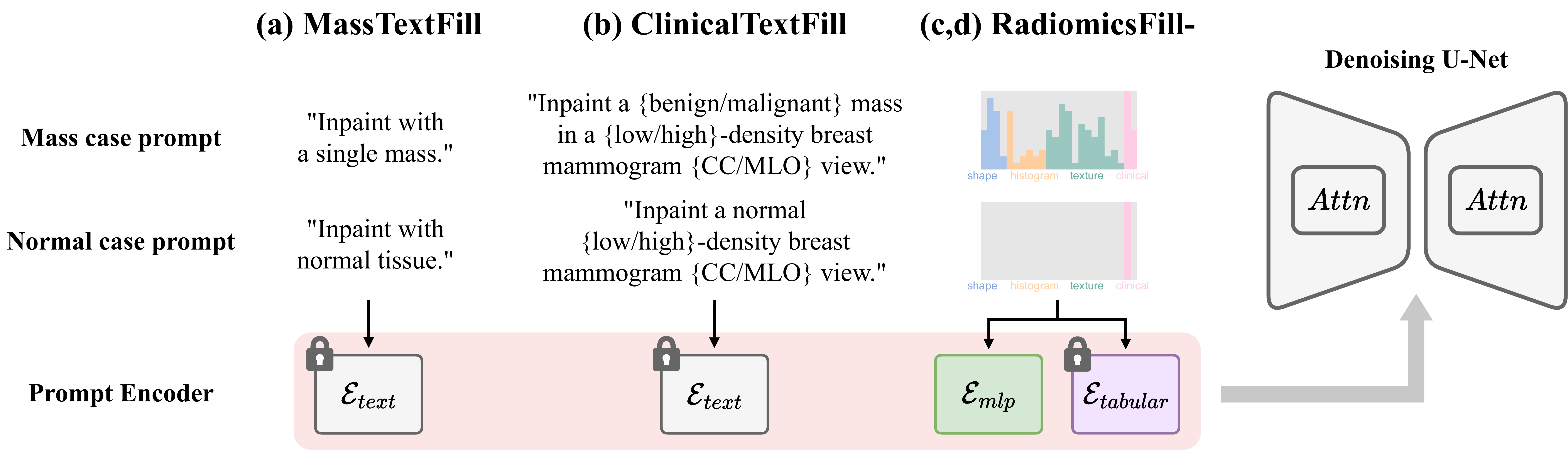}
\caption{\textbf{Different Configurations of Prompt Encoders.} Each configuration employs a prompt encoder specialized for its prompt type. Configurations (a) MassTextFill, (b) ClinicaltextFill, and (d) RadiomicsFill-MET utilize pretrained and frozen prompt encoders, whereas the encoder for (c) RadiomicsFill-MLP consists of three MLP layers and is trained alongside the generator.} \label{fig:main3}
\end{figure}

To assess the performance of \textit{RadiomicsFill-Mammo}, we experimented with four configurations for tumor condition prompts and encoders, as detailed below and illustrated in Fig.~\ref{fig:main3}:

\begin{itemize}
    \item \textbf{(a) MassTextFill:} Simplifies the approach by categorizing inputs into two broad categories: mass and normal cases.
    \item \textbf{(b) ClinicalTextFill:} Utilizes 12 distinct text prompts based on breast density, BI-RADS assessment, and view for more nuanced clinical inputs.
    \item \textbf{(c) RadiomicsFill-MLP} and \textbf{(d) RadiomicsFill-MET:} Both configurations use 67 radiomics features and 2 clinical variables (breast density and BI-RADS) as prompts but differ in their prompt encoder architectures, exploring various encoding strategies.
\end{itemize}

During the denoising U-Net training phase, normal and mass prompts are used to inpaint images, facilitating the learning of distinctions between normal tissue and tumors. For inference, we applied Classifier-Free Guidance (CFG) with tailored prompts to enhance image specificity and clinical relevance \cite{ho2022classifier}. Positive prompts detailed tumor characteristics, while negative prompts ensured differentiation from normal cases. This CFG approach significantly improves the model's accuracy by focusing on tumor-specific attributes and avoiding features typical of normal cases.

\section{Results}

\subsection{How Well Does It Generate Diverse and Realistic Tumors?}

\subsubsection{Radiomics Feature Consistency}
\begin{figure}[t]
\includegraphics[width=\textwidth]{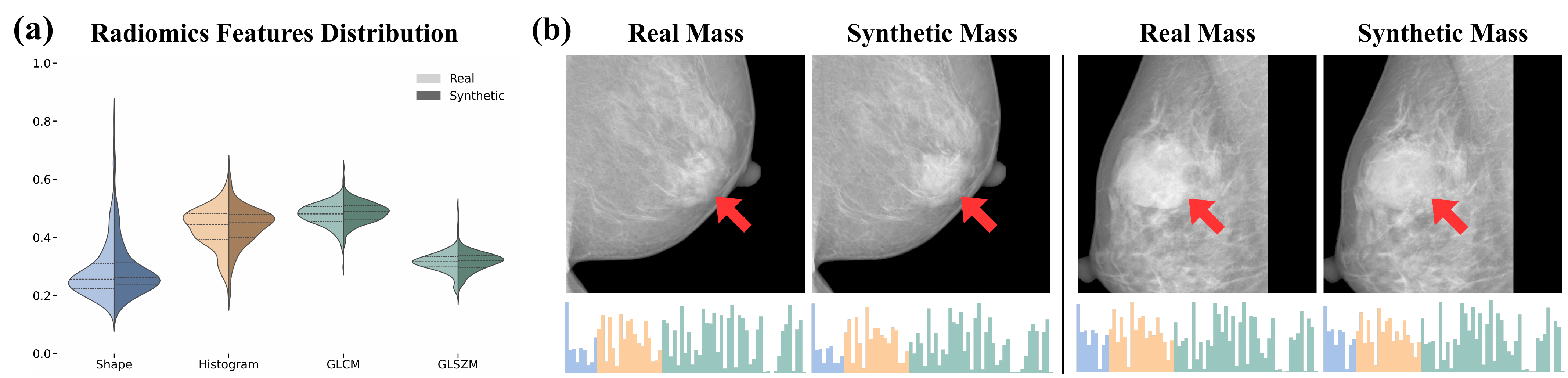}
\caption{ \textbf{Radiomics Feature Consistency Between Real and Synthetic Masses.} (a) Grouped violin plots for shape, histogram, and texture (GLCM and GLSZM) features, contrasting real (left) and synthetic (right) distributions. (b) Visual comparison of real and synthetic masses, with red arrows highlighting masses and histograms visualizing the 67 extracted radiomics features underneath.} \label{fig:main4}
\end{figure}

We assessed \textit{RadiomicsFill}'s ability to generate tumors reflecting specified radiomics features by comparing the radiomics features of generated tumors against those of the input image in unseen test set conditions. We employed a pretrained nnUNet model \cite{isensee2021nnu}, trained on the mass training set, to predict synthetic tumor contours from mass images and bounding box mask images. The high similarity observed in Fig.~\ref{fig:main4} between input and synthetic tumor radiomics features confirms \textit{RadiomicsFill}'s effectiveness in accurately mirroring the given conditions in tumor generation.

\subsubsection{Quality Assessment of Synthetic Tumor }

All models, leveraging tailored prompts, consistently generated high-quality synthetic tumors. The performance was measured with metrics reflecting image fidelity: FID (Fréchet Inception Distance), PSNR (Peak Signal-to-Noise Ratio), SSIM (Structural Similarity Index), and MS-SSIM (Multi-Scale Structural Similarity Index). Lower FID scores indicate better image quality, while higher PSNR, SSIM, and MS-SSIM scores suggest greater similarity to the target images. The strong performance across all prompt types suggests that the stable diffusion backbone is effective as a tumor generator. The results in Table ~\ref{tab:main1} highlight the consistent excellence in synthetic image quality.

\begin{table}[t]
\centering
\caption{\textbf{Synthetic Tumor Quality Assessment.}}\label{tab:main1}
\begin{tabular}{L{3cm} | C{1.8cm} | C{1.8cm} | C{1.8cm} | C{1.8cm}}
\hline
Prompt type & FID ↓ & PSNR ↑ & SSIM ↑ & MS-SSIM ↑\\
\hline
MassTextFill &  0.1144 & 32.4783 &  0.8395 & 0.9461 \\
  \hline
ClinicalTextFill &  0.0989 & \textbf{33.1762} &  0.8430 & 0.9495 \\
  \hline
RadiomicsFill-MLP &  0.0874 & 32.7828 &  0.8322 & 0.9498 \\
  \hline
RadiomicsFill-MET &  \textbf{0.0570} & 32.8009 &  \textbf{0.8535} & \textbf{0.9507} \\
  \hline
\end{tabular}
\end{table}

\subsection{How Can \textit{RadiomicsFill-Mammo} Be Utilized?}
\subsubsection{Enhancing Downstream Tasks}

In evaluating \textit{RadiomicsFill-Mammo}'s impact on mass detection, we generated synthetic data using normal training images with mass prompts. The generated data were used for training a YOLO v8 detection model \cite{yolov8_ultralytics}. We trained models with two regimes: `1x' using only synthetic data, and `2x' combining synthetic data with the actual mass training set. For comparison, the baseline model was trained solely on the real mass training set. To assess mass detection performance, we conducted a comprehensive comparison using commonly employed metrics, with results detailed in Table ~\ref{tab:main2}.

\begin{table}[t]
\centering
\caption{\textbf{Mass detection performance on the VinDr-Mammo test set.} AP; average precision. mAP; mean AP.}\label{tab:main2}
\begin{tabular}{L{3.3cm} | C{1.3cm} | C{1.3cm} | C{1.3cm} | C{1.3cm} | C{1.3cm} | C{1.5cm}}
\hline
Prompt type & L-B AP & L-M AP  & H-B AP  & H-M AP  & m$\textrm{AP}_{50}$  & m$\textrm{AP}_{50-95}$ \\
\hline
Baseline &  0.7500 &  1.0000 &  0.4907 &  0.5056 &  0.5438 &  0.3569 \\
\hline
\hline
MassTextFill-1x &  0.5833 &  0.7500 &  0.4352 &  0.3708 &  0.4378 &  0.2488 \\
MassTextFill-2x &  0.7778 &  0.8333 &  0.4815 &  0.4045 &  0.4900 &  0.3301 \\
\hline
ClinicalTextFill-1x &  0.5417 &  0.6667 &  0.4047 &  0.2921 &  0.3871 &  0.2245 \\
ClinicalTextFill-2x &  0.9444 &  1.0000 &  0.5648 &  0.5281 &  0.6006 &  0.3754 \\
\hline
RadiomicsFill-MLP-1x &  0.5833 &  0.8333 &  0.2963 &  0.3708 &  0.3733 &  0.2227 \\
RadiomicsFill-MLP-2x &  0.8333 &  0.9167 &  0.5648 &  0.5730 &  0.6037 &  0.3784 \\
\hline
RadiomicsFill-MET-1x &  0.5000 &  0.7500 &  0.3796 &  0.4157 &  0.4240 &  0.2550 \\
RadiomicsFill-MET-2x &  \textbf{1.0000} &  \textbf{1.0000} &  \textbf{0.5648} &  \textbf{0.6292} &  \textbf{0.6429} &  \textbf{0.3935} \\
\hline
\end{tabular}
\end{table}

\subsubsection{Visualizing Condition-Specific Tumor Variations}
\begin{figure}[t]
\includegraphics[width=\textwidth]{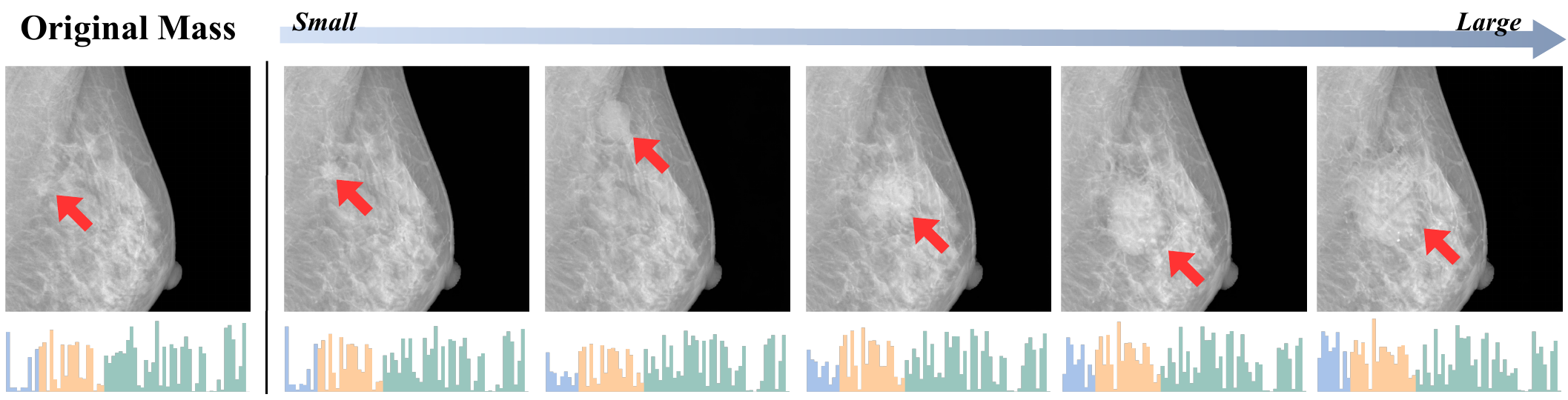}

\caption{\textbf{Progressive Shape Variation in Synthetic Tumors}. Starting with an image containing a mass, we use a normal prompt to replace the mass area with normal tissue. Then, at random positions, we generate synthetic tumors of increasing size, keeping histogram and texture features consistent. Red arrows in each image point to the mass. Below each image, histograms display the radiomics features extracted from the tumor areas.}

\label{fig:main5}
\end{figure}

Fig.~\ref{fig:main5} showcases synthetic tumors generated with progressively increasing the shape feature of size, while other features remain constant. This visualization demonstrates our model's ability to adapt to specific tumor characteristics, illustrating the potential for generating diverse, tailored synthetic tumors based on specific condition prompts.

\subsubsection{Fine-Tuning with External Validation Dataset}
This section underscores \textit{RadiomicsFill-Mammo}'s capability to adapt to external datasets, using the INbreast dataset \cite{moreira2012inbreast} from Portugal as a case study. To address data scarcity and enhance mass detection, we specifically fine-tuned the denoising U-Net of \textit{RadiomicsFill-Mammo} by training only the added Low-Rank Adaption (LoRa) module \cite{hu2021lora}. This focused fine-tuning strategy demonstrates an effective adaptation even with limited data availability. To enrich the dataset, we generated synthetic masses on INbreast's normal images using mass prompts from the VinDr-Mammo trainset, generating an augmented dataset for training the mass detection model. Due to the limited size of the INbreast dataset, we proceeded directly to test set evaluation after 300 epochs of balanced training, without employing a separate validation set. The training and test sets included 7 L-B, 30 L-M, 4 H-B, and 3 H-M masses for training, and 9 L-B, 34 L-M, 18 H-B, and 6 H-M masses for testing, respectively using stratified sampling at the patient level. The improvements in diagnostic accuracy achieved through this refined fine-tuning method, as evidenced in Table ~\ref{tab:main3}, highlight the enhanced utility of \textit{RadiomicsFill-Mammo} in real-world scenarios, even with limited available data.

\begin{table}[t]
\centering
\caption{\textbf{Mass detection performance on the INbreast dataset.}}\label{tab:main3}
\begin{tabular}{L{3.3cm} | C{1.3cm} | C{1.3cm} | C{1.3cm} | C{1.3cm} | C{1.3cm} | C{1.5cm}}
\hline
Prompt type & L-B AP & L-M AP  & H-B AP  & H-M AP  & m$\textrm{AP}_{50}$  & m$\textrm{AP}_{50-95}$ \\
\hline
Baseline &  0.3333 &  0.7778 &  0.3846 &  0.2500 &  0.5800 &  0.3978 \\
\hline
\hline
MassTextFill &  0.5000 &  0.7778 &  0.6923 &  0.2500 &  0.6800 &  0.4511 \\
\hline
ClinicalTextFill &  0.5000 &  0.7778 &  0.6154 &  0.5000 &  0.6800 &  0.4533 \\
\hline
RadiomicsFill-MLP &  \textbf{0.6667} &  \textbf{0.8519} &  0.4615 &  0.5000 &  0.7000 &  0.4689 \\
\hline
RadiomicsFill-MET &  0.5000 &  0.8148 &  \textbf{0.7692} &  \textbf{0.5000} &  \textbf{0.7400} &  \textbf{0.4711} \\
\hline
\end{tabular}
\end{table}

\section{Conclusion}

Our study introduces the pioneering application of utilizing radiomics features for generating synthetic tumors. We have successfully validated the generator's precision in mirroring specified tumor conditions, markedly enhancing mass detection capabilities when used as a data augmentor. This improvement is especially significant for high-density masses, which are often difficult to detect due to their reduced visibility in dense breast tissue. The findings demonstrate that defining specific tumor conditions in conjunction with an appropriately configured prompt encoder enables the generator to produce a wide range of synthetic tumors, preventing the generation of typical results. The innovative framework of \textit{RadiomicsFill-Mammo}, which incorporates opposite breast images and clinical variables such as breast density and BI-RADS assessment as additional conditions for tumor generation, establishes a clinically meaningful approach. As the first endeavor in the \textit{RadiomicsFill} series, \textit{RadiomicsFill-Mammo} unveils substantial potential to advance the field of medical imaging research, achieving significant outcomes. The ability of \textit{RadiomicsFill} to generate tumors with specific attributes highlights its utility for a range of clinical purposes, including simulating treatment responses. Furthermore, this work paves the way for applying \textit{RadiomicsFill} techniques across different medical imaging fields, setting a promising direction for our future research.

\begin{credits}
\subsubsection{\ackname} This study was supported by National Research Foundation (NRF-2020M3E5D2A01084892), Institute for Basic Science (IBS-R015-D1), AI Graduate School Support Program(Sungkyunkwan University) (RS-2019-II190421), ICT Creative Consilience program (RS-2020-II201821), and the Artificial Intelligence Innovation Hub program (RS-2021-II212068).

\subsubsection{\discintname}
The authors have no competing interests to declare that are relevant to the content of this article.
\end{credits}
%
%
%
\bibliographystyle{splncs04}
\bibliography{ref}

\newpage
\appendix
\renewcommand\thefigure{\Alph{figure}}    
\setcounter{figure}{0}  
\section*{Supplementary Material}

\begin{figure}
\includegraphics[width=\textwidth]{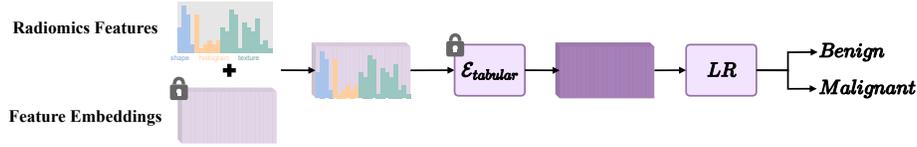}
\caption{\textbf{Evaluating Framework for the Efficiency of the Pretrained Tabular Encoder.} During the training phase, clinical variables were included alongside radiomics features. For evaluation, all clinical variables were masked, and only the 67 radiomics features were fed into the encoder. The encoder's output, once flattened, served as input for a logistic regression model tasked with a binary classification of BI-RADS into benign and malignant categories. The predictive performance of this model highlights the efficacy of MET in capturing clinically relevant information from radiomics features alone.} \label{fig:appendixA}
\end{figure}

\begin{figure}
\centering
\includegraphics[width=0.8\textwidth]{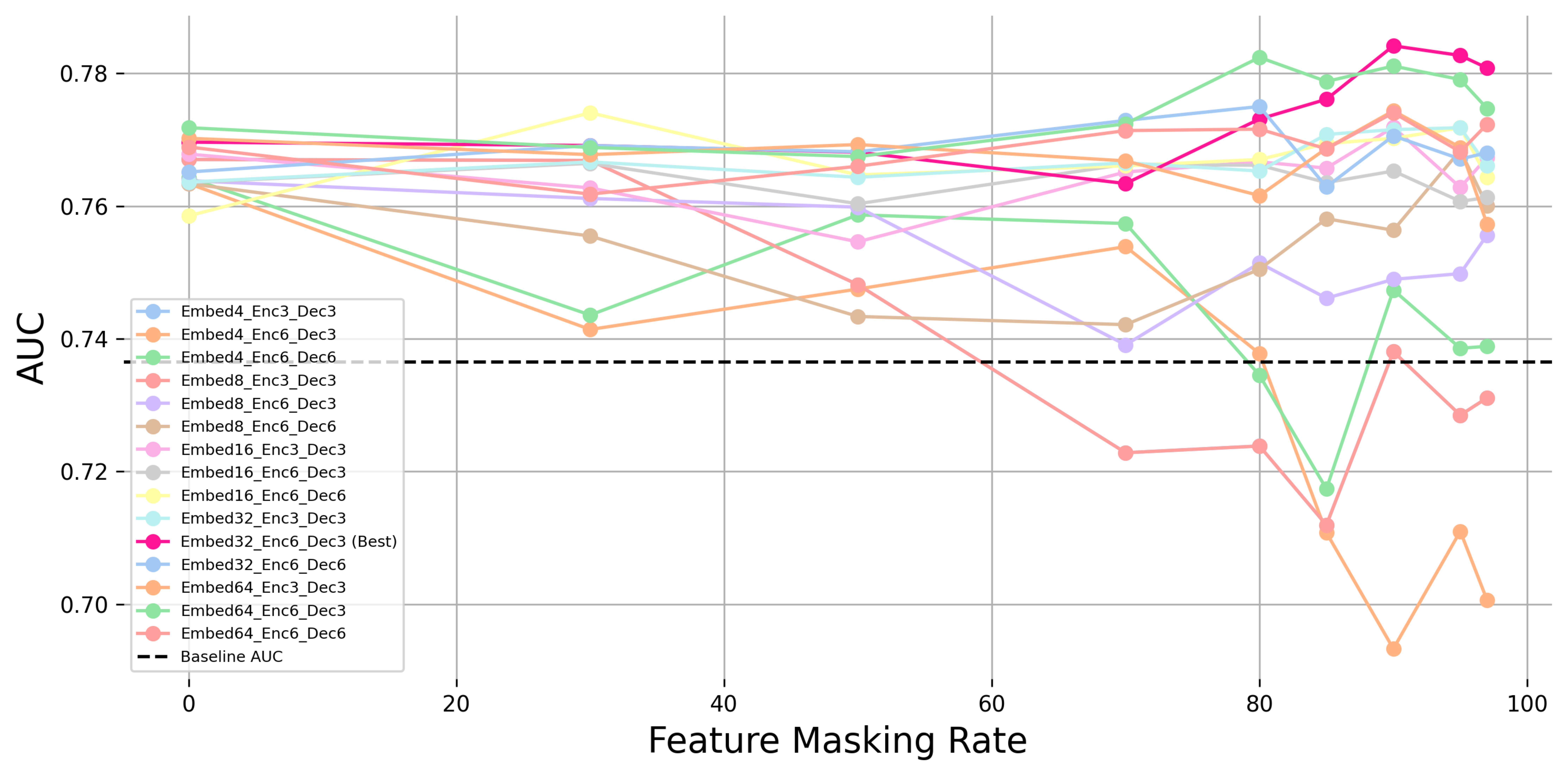}
\caption{\textbf{Optimizing MET Parameters: AUC Performance Across Masking Rates.} This graph tracks AUC scores for benign/malignant predictions at various feature masking rates to determine MET's optimal parameters. Highlighted in hot pink, the best-performing setup—featuring a 32-dimensional feature embedding, 6 encoder layers, and 3 decoder layers at a 90\% masking rate—was chosen as the prompt encoder configuration for RadiomicsFill-MET.} \label{fig:appendixB}
\end{figure}

\begin{figure}
\includegraphics[width=\textwidth]{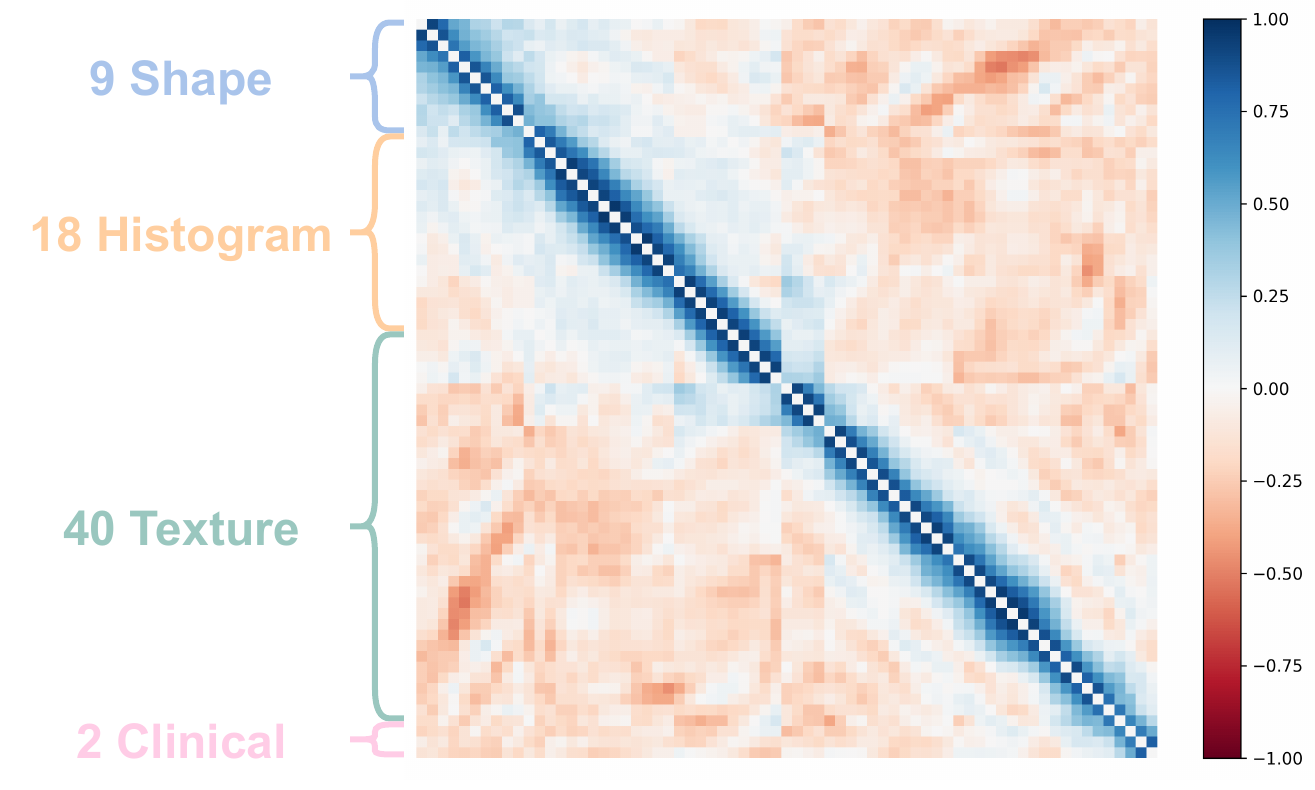}
\caption{\textbf{Cosine Similarity between MET Feature Embeddings.} This heatmap visualizes the cosine similarity for MET's 69 feature-specific embeddings, highlighting strong intra-group correlations within shape, histogram, and texture clusters. The clear block diagonal structure observed signifies the encoder's proficiency in grouping and differentiating feature categories.}  \label{fig:appendixC}
\end{figure}

\end{document}